\documentclass[pra,reprint,superscriptaddress,floatfix,aps]{revtex4-2}
\usepackage{amsmath}
\usepackage{amsfonts}
\usepackage{graphicx}
\usepackage{physics}
\usepackage[usenames]{color}
\usepackage[colorlinks,linktocpage,bookmarks=false,citecolor=blue,linkcolor=red,urlcolor=blue]{hyperref}
\usepackage{hyperref}
\usepackage{cleveref}

\begin{document}

\title{Non-equilibrium dynamics of the disordered {\em Power of Two} model}
\author{Kunal Singh}
\author{Sayan Choudhury}
\email{sayanchoudhury@hri.res.in}
\affiliation{Harish-Chandra Research Institute, a CI of Homi Bhabha National Institute, Chhatnag Road, Jhunsi, Allahabad 211019}

\date{\today}

\begin{abstract}
Motivated by recent experimental realizations of programmable spin models with long-range interactions, we investigate the non-equilibrium dynamics of the \emph{Power-of-Two} (PWR2) model. This model consists of sparse long-range couplings between spin-$1/2$ objects separated by $d = 2^n$. In the absence of disorder, the system exhibits rapid scrambling and fast thermalization. We explore the impact of disorder in this system by analyzing the time evolution of the survival probability, half-chain entanglement entropy, and out-of-time-ordered correlators (OTOCs). We find that sufficiently strong disorder suppresses information spreading and induces localization. Remarkably, in the strong-disorder regime, the OTOCs display a non-monotonic spatial profile arising from the intrinsic nonlocality of the interactions, signaling qualitatively distinct scrambling dynamics compared to conventional long-range interacting systems. To characterize the localization transition, we extract the critical disorder strength $h_c$ from the spectral statistics and the eigenstate entanglement. We observe that $h_c$ increases with system size. Furthermore, at a fixed disorder strength, the eigenstate-averaged entanglement entropy increases with system size, while the inverse participation ratio decreases, indicating enhanced delocalization at larger sizes. These results collectively suggest that the PWR2 model remains ergodic in the thermodynamic limit for any finite disorder strength.

\end{abstract}
	
\maketitle
\section{Introduction} 
The far-from-equilibrium dynamics of isolated many-body systems have received significant attention in recent years~\cite{nandkishore2015many,abanin2019colloquium,altman2018many,mori2018thermalization,mallayya2019prethermalization,fisher2023random,vallejo2025single}. Rapid advances in the capabilities of various experimental platforms such as Rydberg atom arrays~\cite{bluvstein2021controlling,liang2025observation,manovitz2025quantum}, trapped ions~\cite{smith2016many,morong2021observation}, and superconducting qubit processors~\cite{andersen2025thermalization,braumuller2022probing,mi2021information} have propelled this. On the one hand, this has led to the development of quantum technologies, where non-equilibrium protocols play a crucial role~\cite{colombo2022time,moon2026sensing}. On the other hand, it has led to important insights about fundamental questions regarding the nature of quantum thermalization~\cite{kaufman2016quantum,deutsch2018eigenstate} and non-equilibrium phases of matter~\cite{zaletel2023colloquium,else2020discrete}. A particularly intriguing direction of research has been to understand the role of information scrambling and its connections to chaos and thermalization~\cite{lewis2019dynamics,xu2024scrambling}. Information scrambling describes the process by which locally encoded quantum information is spread by interactions into non-local degrees of freedom~\cite{swingle2016measuring,landsman2019verified,bhattacharyya2022quantum}. Consequently, this information becomes inaccessible to local probes~\cite{xu2019locality,lewis2019unifying}. Interestingly, scrambling can be harnessed for several quantum information protocols such as time-reversed metrology~\cite{li2023improving,zheng2024improving,ge2025information}, many-body teleportation~\cite{schuster2022many}, and quantum error correction~\cite{choi2020quantum}.\\
Black holes are believed to be the fastest scramblers in nature~\cite{sekino2008fast,lashkari2013towards}. Interestingly, in recent years, several fast-scrambling many-body models have been proposed that have profound connections to quantum black holes. Initial investigations in this direction primarily focused on the celebrated Sachdev-Ye-Kitaev model~\cite{sachdev1992gapless,sachdev2010holographic,kitaev2015simple}, which features random infinite-range interactions between Majorana fermions~\cite{maldacena2016remarks}. Notably, recent investigations have revealed that clean chaotic systems can also exhibit fast scrambling~\cite{bentsen2019treelike,belyansky2020minimal,li2020fast,kuno2023phase,kuriyattil2023onset,hashizume2021deterministic,toga2025fast,li2022fast}. A particularly intriguing example of such a clean fast scrambler is the {\emph {`Power of Two'}} (PWR2) model~\cite{bentsen2019treelike}, which is characterized by sparse long-range interactions between spin-1/2 particles. This model has been realized in cold-atom experiments~\cite{periwal2021programmable} and can serve as a platform to generate metrologically useful entangled states~\cite{kuriyattil2025entangled} and investigate geometry-driven phase transitions~\cite{gunning2025geometry}.\\
In this work, we investigate the effect of on-site disorder on the quench dynamics, spectral statistics, and eigenstate properties of the PWR2 model. We find that the growth of the out-of-time-ordered correlators (OTOCs) and entanglement entropy is considerably suppressed in the presence of disorder. This is accompanied by an increase in survival probability, thereby signaling the persistence of the memory of the initial state and consequently non-ergodic behavior. Notably, we find that the OTOCs exhibit a non-monotonic spatial dependence due to the non-local nature of couplings in this system. We analyze the spectral statistics of this model and find that the system undergoes a transition from chaotic to many-body localized state when the disorder strength $h$ exceeds a critical value, $h_c$, where $h_c$ diverges with the system size, $L$. We find analogous results in the behavior of the eigenstate entanglement and the inverse participation ratio, and conclude that the system does not host a many-body localized phase in the thermodynamic limit. Our analysis reveals that the interplay of interactions and disorder can lead to rich non-equilibrium behavior in the PWR2 model.\\
\begin{figure*}[ht]
   \includegraphics[width=0.9\textwidth]{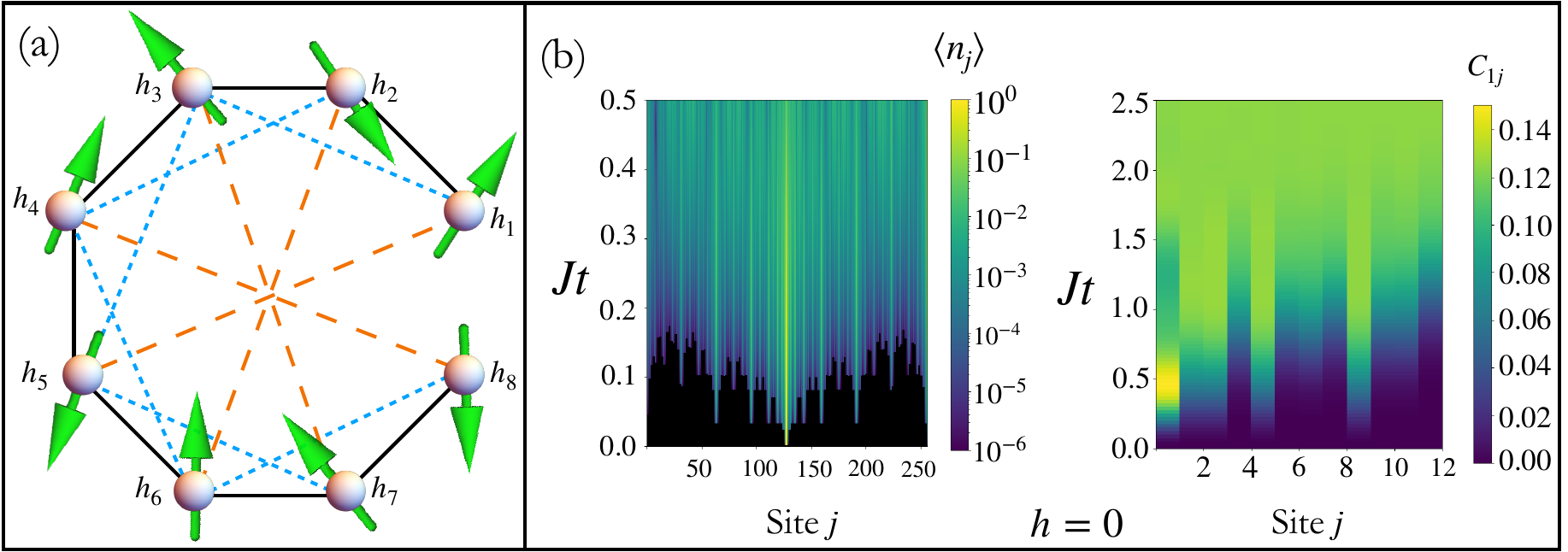}
\caption{(a) Schematic illustration of the {\emph {Power of Two}} model with open boundary conditions. The interaction graph couples spins on sites separated by a power of two (see Eqns.~\ref{eq:Ham} and ~\ref{eq:Ham2}). (b) The dynamics of the PWR2 model in the disorder-free limit ($h=0$). The left panel shows the time-evolution of the occupation, $\langle n_j \rangle = \langle S_j^{+} S_j^{-} \rangle$ for the single-magnon state $\vert \psi \rangle_{\rm SM}$ (Eq.~\ref{eq:SM-state}). The right panel shows the time-evolution of the OTOC, $C_{1j}(t)$ (Eq.~\ref{eq:OTOC}) in the zero-magnetization sector. These results demonstrate that this system exhibits fast scrambling and consequent rapid thermalization.}
    \label{fig:Figure1}
\end{figure*}
This paper is organized as follows. We introduce the model in Sec.~\ref{sec:Model} and explore the quench dynamics of the system in Sec.~\ref{sec:Quench}. We analyze the spectral statistics and eigenstate properties of the system in Sec.~\ref{sec:Levelstat}. Finally, we conclude this work with a summary of results and outlook for future research direction in Sec.~\ref{sec:Conclusion}.

\section{Model}
\label{sec:Model}

We investigate the quench dynamics of the PWR2 model first introduced in Ref.~\cite{bentsen2019treelike} in the presence of on-site disorder:\\
\begin{equation}
    H_{\rm PWR2}=\sum_{\substack{i,j=1 \\ i \ne j}}^N J_{ij} \left(S_i^{\rm x} S_j^{\rm x} +S_i^{\rm y} S_j^{\rm y} \right)+\sum_i h_iS_i^{\rm z}
    \label{eq:Ham}
\end{equation}
where $S_i^{\alpha}=\frac{1}{2}\sigma_i^{\alpha}$, $\sigma$ are the standard Pauli matrices, and $h_i$ are the random magnetic fields such that $h_i \in [-h,h]$ and uniformly distributed. The crucial feature that distinguishes the PWR2 model from other long-range interacting models is that the spins on site $i$ and $j$ interact only when $|i-j|$ is a power-of-two:
\begin{equation}
    J_{ij} =
\begin{cases} 
J & \text{if } |i-j|=1,2,4,8 \ldots \\ 
0 & \text{otherwise } 
\end{cases}
\label{eq:Ham2}
\end{equation} 
A schematic illustration of this sparse-coupling interaction graph is provided in Fig.~\ref{fig:Figure1}(a). We note that $H_{\rm PWR2}$ commutes with the total z-magnetization $M^z=\sum_{i=1}^N S_i^z$ and we analyze the dynamics in the single-magnon ($M^z = (1-N/2)$) and the zero-magnetization ($M^z = 0$) sectors in the following analysis. \\
Before proceeding to study the effect of disorder, it is instructive to review the dynamical properties of this model in the disorder-free limit. In this regime, this system represents a transition between Archimedean and ultrametric geometries, and consequently, the system can exhibit fast scrambling. This can be seen by first examining the time-evolution of the system for the 1-magnon initial state:
\begin{equation}
\vert \psi(t=0) \rangle = \vert \psi\rangle_{\rm SM} = S^{+}_{N/2} \vert \downarrow \downarrow \downarrow \ldots \downarrow \downarrow \rangle
\label{eq:SM-state}
\end{equation}
where $S^{+} = S^x + i S^y$, and $\vert \downarrow \downarrow \downarrow \ldots \downarrow \downarrow \rangle$ represents the $-z$-polarized initial state. We characterize information spreading for this initial state by computing the magnon occupation at site $j$:
\begin{equation}
\langle n_j \rangle = (1+2 \langle S_j^z\rangle)/2.
\label{eq:magnon}
\end{equation}
Our results (shown in Fig.~\ref{fig:Figure1}(b)) clearly demonstrate that the magnon excitation spreads rapidly over the entire system in a time, $t \propto \log(N)$.\\
To gain further insights into the non-equilibrium behavior of this system, we investigate information scrambling in the zero-magnetization sector by computing the OTOC~\cite{swingle2018unscrambling,hosur2016chaos}:
\begin{equation}
C_{ij}(t)= \left\langle \left[ S^z_i(t),S^z_j(0)\right]^\dagger\left[S^z_i(t),S^z_j(0)\right] \right \rangle
\label{eq:OTOC}
\end{equation}
where the angular brackets denote the infinite-temperature average. Fast scrambling systems are characterized by an exponential growth of $C_{ij}\sim \exp(\lambda T)/N^{\alpha}$, where $\lambda$ is the Lyapunov exponent~\cite{maldacena2016bound,bentsen2019fast}. As shown in Fig.~\ref{fig:Figure1}(b), $C_{ij}$ grows super-ballistically, thereby signaling fast scrambling; the scrambling time, $t^{\ast}$ when $C$ approaches its saturation value, $C_{\rm chaotic}$, scales logarithmically with $N$, $t^{\ast} \propto \log(N)/\lambda$~\cite{bentsen2019treelike}. We now proceed to examine the dynamics of this system in the presence of disorder.

\begin{figure}[ht] 
    \includegraphics[width=\linewidth]{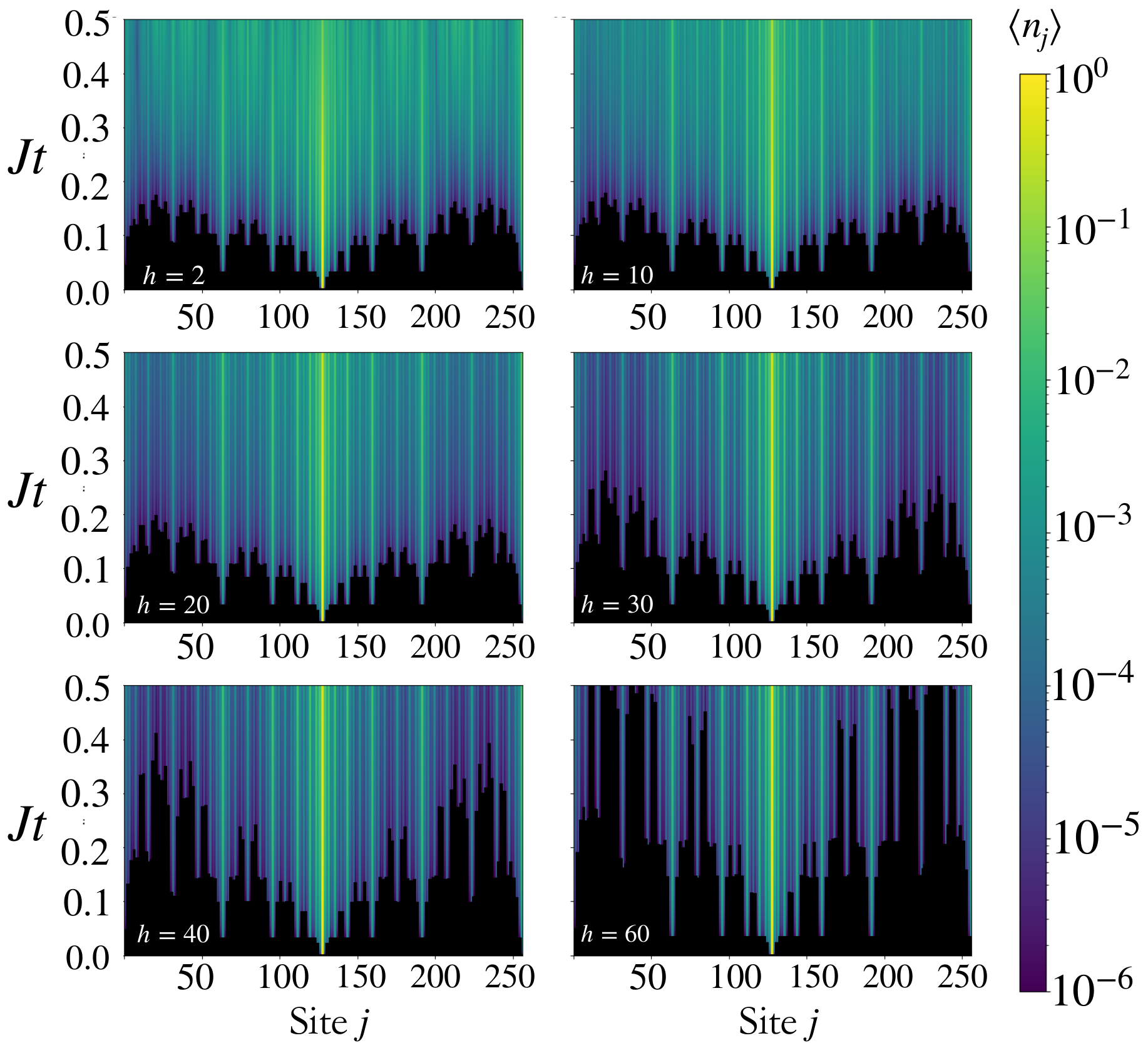} 
    \caption{The time-evolution of the 1-magnon initial state, $\vert \psi \rangle_{\rm SM}$ (Eq.~\ref{eq:SM-state}) for N=256 for various values of the disorder strength, $h$. While, increasing disorder strength leads to greater localization, the spread of the magnon excitation does not follow a light-cone. Instead at strong disorder, the excitation number  $\langle n_j \rangle$ exhibits a  non-monotonic dependence on $\vert j - L/2 \vert$ due to the non-local nature of the model. The results have been obtained from disorder-averaging over 1000 realizations.}
    \label{fig:Figure2}
\end{figure}
\section{Quench Dynamics}
\label{sec:Quench}

Strong disorder induces non-ergodicity and possibly many-body localization (MBL) in chaotic short-range interacting quantum systems~\cite{pal2010many,alet2018many,sierant2025many}. This non-ergodic behavior is usually probed experimentally by analyzing the initial state memory retention~\cite{schreiber2015observation,choi2016exploring}, and the logarithmic growth of entanglement entropy~\cite{lukin2019probing} and OTOCs~\cite{hayata2025digital} after a quantum quench. Notably, MBL is purported to be an eigenstate phase characterized by emergent integrability and a consequent strong violation of the Eigenstate Thermalization Hypothesis~\cite{huse2014phenomenology}. We note that while MBL has been most extensively studied for short-range models, certain classes of long-range interacting systems may also host the MBL phase~\cite{nandkishore2017many,nag2019many,sierant2019many,prasad2021many}. A natural question in this context is whether the PWR2 model hosts the MBL phase. In this section, we partially address this issue by analyzing the dynamics of the system after a quantum quench.\\
We start our analysis by examining the dynamics of this system in the single-magnon sector. Following the procedure outlined in Sec.~\ref{sec:Model}, we compute the magnon occupation for the $\vert \psi(t=0) \rangle_{\rm SM}$ initial state for different values of the disorder strength, $h$. Our results are shown in Fig.~\ref{fig:Figure2}. We find that for weak disorder strength, the initially localized spin excitation spreads rapidly through the system, in a manner analogous to the disorder-free case. However, with increasing disorder strength, the excitation gets increasingly localized. Interestingly, due to the non-local nature of the couplings, a `light cone' never appears in the dynamics, and the magnon excitation $\langle n_j(t) \rangle$ spreads non-locally, thereby leading to a non-monotonic dependence on the distance of the site $j$ from the center of the chain ($|j-L/2|$). This feature distinguishes the PWR2 model from other long-range interacting disordered systems.\\
\begin{figure}[t]
\includegraphics[width=\linewidth]{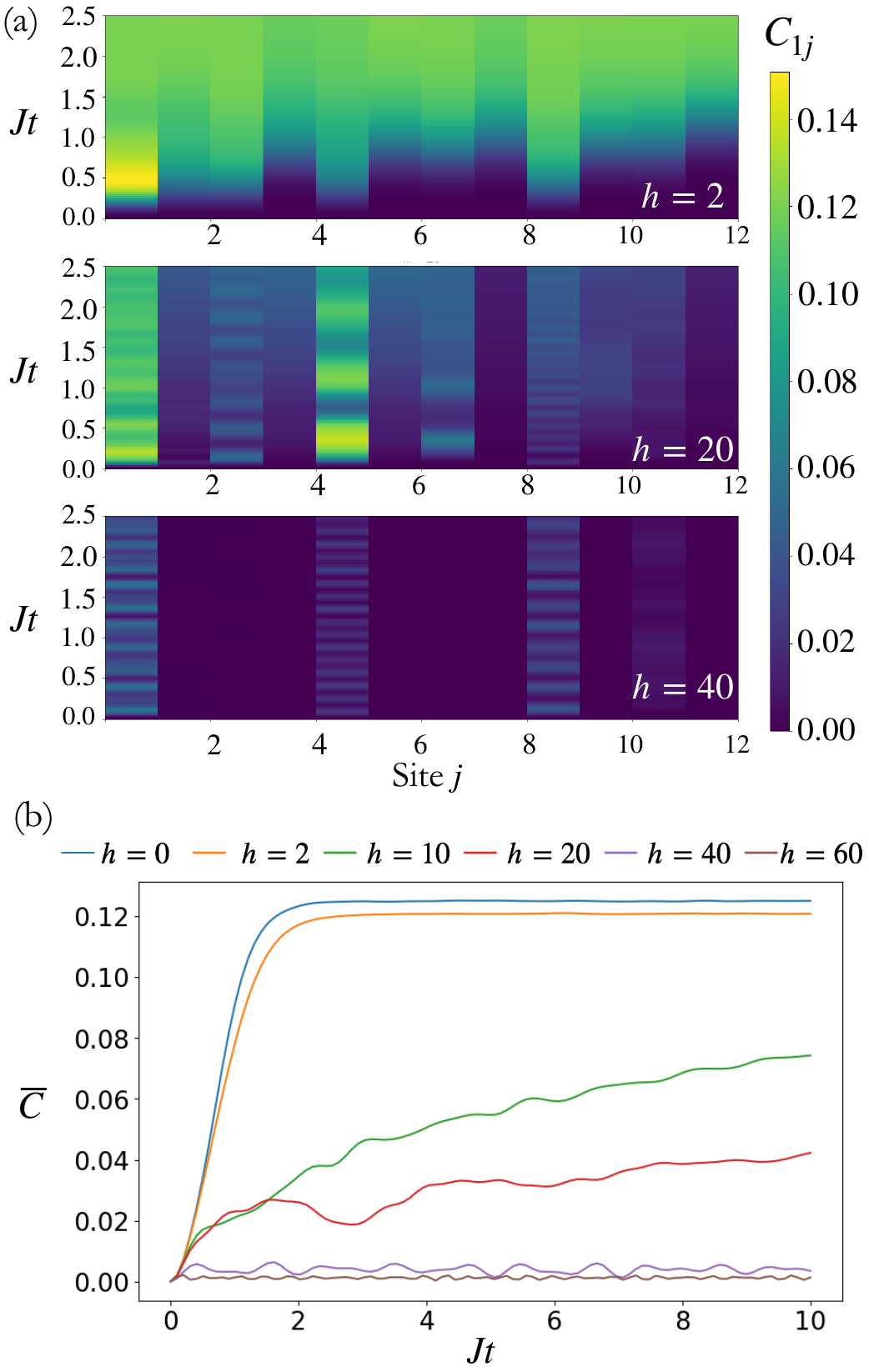} 
\caption{(a) The OTOC, $C_{1j}$ for 3 representative values of disorder strength ranging from weak to strong disorder. In the weak disorder regime, the OTOC spreads super-ballistically. However, in the presence of stronger disorder, the growth slows down. Furthermore, $C_{1j}$ exhibits a non-monotonic dependence on $j$ due to the non-local nature of the interactions. (b) The time-evolution of the spatially averaged OTOC, $\overline{C}(t)$ (Eq.~\ref{eq:AvgOTOC}). In the weak-disorder regime, $\overline{C}(t)$ saturates to $C_{\rm chaotic} \sim 0.125$ within a very short time. Increasing the disorder strength leads to a slower growth of $\overline{C}(t)$, and in the strong disorder regime $\overline{C}(t) \sim 0$. These results have been obtained for $N=12$ by averaging over 100 disorder realizations.}
    \label{fig:Figure3}
\end{figure}
We now proceed to examine information scrambling in the zero-magnetization sector by computing the OTOC, $C_{ij}(t)$ (defined in Eq.~\ref{eq:OTOC}) for various values of the disorder strength, $h$. Our results are reported in Fig.~\ref{fig:Figure3}. Analogous to the single-magnon dynamics discussed above, $C_{ij}(t)$ grows super-ballistically in the weak disorder regime, and exhibits greater localization in the presence of stronger disorder. Intriguingly, in the strong disorder regime, $C_{ij}(t)$ exhibits a strong non-monotonic dependence on the disorder strength, $h$, due to the non-local nature of the interactions. We further characterize the scrambling in this system by computing the spatially averaged OTOC:
\begin{equation}
\overline{C}(t) =\frac{1}{N-1}\sum_{j>1}C_{1j}(t).
\label{eq:AvgOTOC}
\end{equation}
Our results shown in Fig.~\ref{fig:Figure3} clearly demonstrates that $\overline{C}(t)$ grows rapidly and saturates to $C_{\rm chaotic} \sim 0.125$ in the weak disorder regime. Furthermore, the greater localization induced by increasing disorder leads to a slower growth of $\overline{C}(t)$. Notably, in the strong disorder regime, $\overline{C}(t)$ saturates to a very low value $(\sim 0)$.\\
\begin{figure}[t]
    \includegraphics[width=\linewidth]{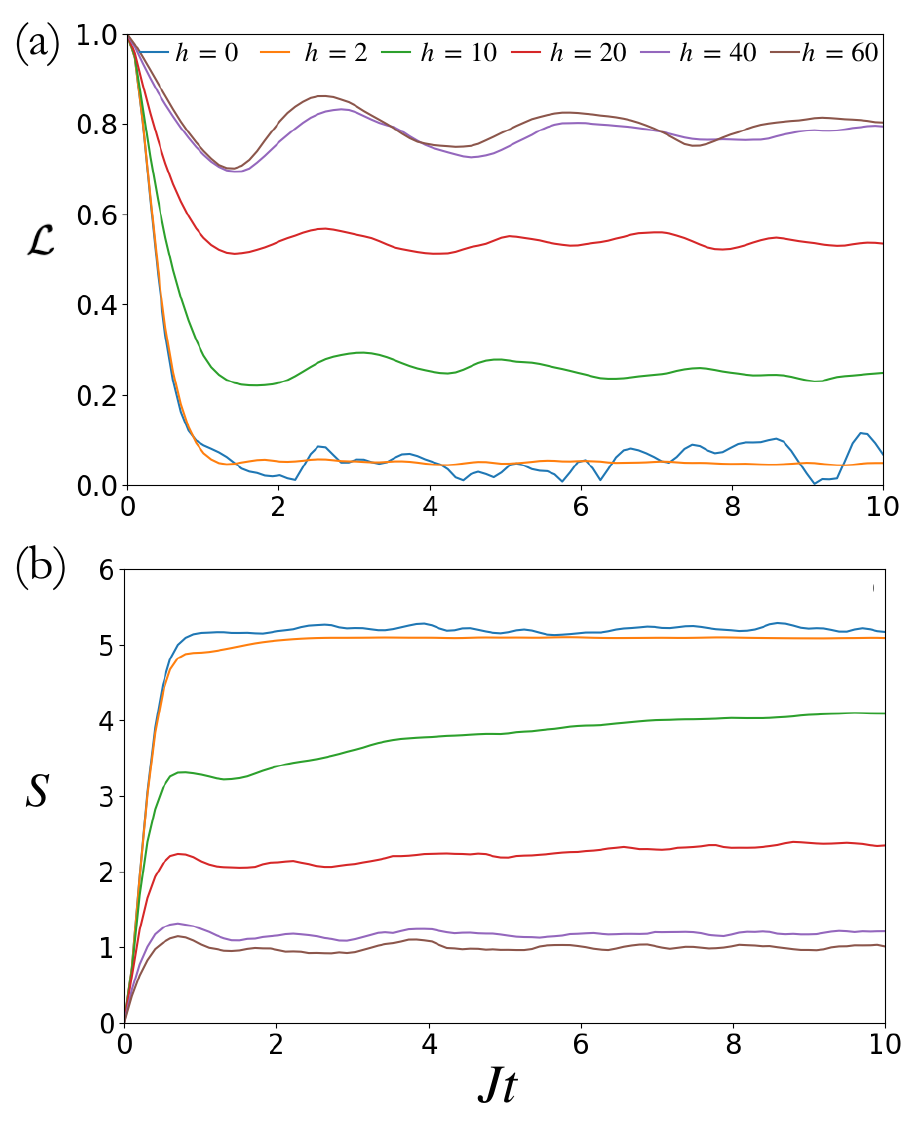} 
    \caption{(a) The time-evolution of the survival probability, $\mathcal{L}(t)$ (Eq.~\ref{eq:Fidelity}) for different values of $h$. In the weak disorder regime, $\mathcal{L}(t)$ quickly decays to zero due to the fast information scrambling. Increasing disorder leads to a slower decay and a higher saturation value of $\mathcal{L}(t)$.  (b) The growth of the half-chain entanglement entropy, $S(t)$ for various value of the disorder strength, $h$. In the weak disorder regime, $S(t)$ quickly grows and saturates to its maximum value. Increasing disorder suppresses the rate of growth and the saturation value of $S$. These results have been obtained for the domain wall initial state, $\vert \psi \rangle_{\rm DW}$; $N$ has been set to 12, and the disorder averaging has been performed for 100 disorder realizations.}
    \label{fig:Figure4}
\end{figure}
We gain further insights into the disorder-induced non-ergodicity by examining the time-evolution of the system from a domain-wall initial state:
\begin{equation}
|\psi(t=0)\rangle = \vert \psi \rangle_{\rm DW} = \left | \uparrow \uparrow \ldots \uparrow \uparrow\downarrow \downarrow \ldots\downarrow\downarrow  \right \rangle,
\end{equation} 
where $\vert \uparrow \rangle_i (\vert \downarrow\rangle_i)$ is an eigenstate of $\sigma_i^z$ with eigenvalue $+1(-1)$. We compute the time-evolution of the survival probability:
\begin{equation}
\mathcal{L}(t)=|\langle\psi(t)|\psi(0)\rangle|^2,
\label{eq:Fidelity}
\end{equation} 
and the half-chain entanglement entropy:
\begin{equation}
S(t)=-{\rm Tr}\left( \rho_{L} (t)\log_2(\rho_{L}(t))\right),
\label{eq:Entanglement}
\end{equation}
where $\rho_L = {\text Tr_R} (\vert \psi \rangle \langle \psi \vert)$ is the reduced density matrix, obtained by tracing over the degrees of freedom of one half of the chain. We note that in the weak disorder regime, $\mathcal{L}(t)$ decreases rapidly to $0$ indicating a loss of the memory of the initial conditions. This is accompanied by a rapid increase of $S$ to its steady state value, $S_{\rm fin}(h)$; for a fixed system size, $N$, $S_{\rm fin} (h)$ is maximum for $h=0$.  Increasing disorder leads to a slower decrease (increase) of $\mathcal{L}(t)\,\, (S(t))$; crucially, $\mathcal{L}(t) \,\, (S(t))$ saturates to a higher (lower) value with stronger disorder, thereby indicating greater localization. Our results are shown in Fig.~\ref{fig:Figure4}. The results from our analysis indicate that strong disorder induces localization in the PWR2 model. We note however that the quench dynamics alone is insufficient to establish whether this system exhibits many-body localization. To explore this issue further, we now proceed to examine the spectral statistics and eigenstate properties of this model.

\section{Spectral Statistics and Eigenstate Properties}
\label{sec:Levelstat}
\begin{figure}[ht]
\includegraphics[width=\linewidth]{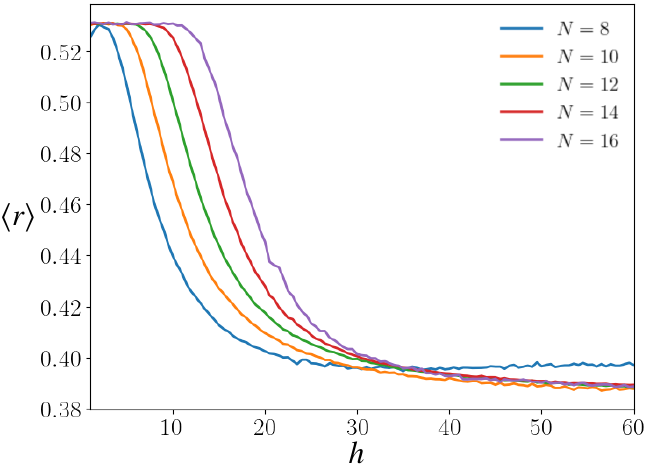} 
\caption{The adjacent gap ratio, $\langle r \rangle$ as a function of disorder strength $h$. The results have been averaged over 60000 realizations for $N=8$, 20000 realizations for $N=10$ and $12$, and 200 realizations for $N=14$ and $16$ respectively. These results have been obtained for the  middle 1/3 spectra. Increasing $h$ leads to a transition of the spectral statistics from Wigner-Dyson to Poisson, thereby indicating localization. The $\langle r \rangle$ curve shifts towards the right with increasing $N$ indicating that the critical disorder strength for localization, $h_c \rightarrow \infty$ in the thermodynamic limit.} 
\label{fig:Figure5}
\end{figure}

\begin{figure}[t]
\includegraphics[width=\linewidth]{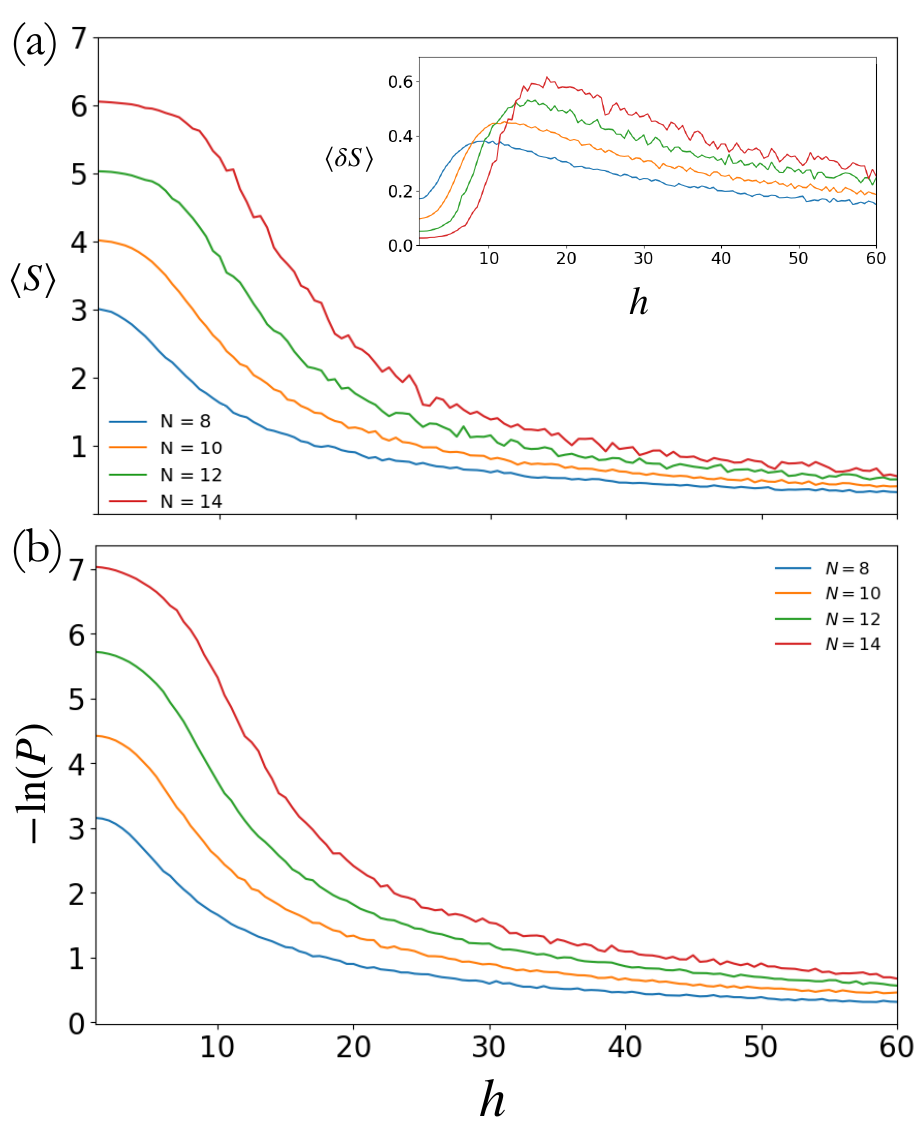} 
        \caption{(a) The half-chain entanglement entropy, averaged over the central 20 eigenstates, $\langle S \rangle$, as a function of disorder strength, $h$. In the low disorder regime, $\langle S \rangle$ is very close to the Page value. Increasing disorder leads to a smaller value of $S$. However, $\langle S \rangle$ always increases with $L$ for a fixed value of $h$. The inset shows the variance of the eigenstate entanglement, $\delta S$. The peak of $\delta S$ corresponds to the thermalization-MBL transition point, and it shifts to the right with increasing $L$. The disorder-averaging has been performed over  1000, 600, 200 and 100 disorder realizations for $N=8$, $10$, $12$ and $14$ respectively. (b) The natural log of inverse participation ratio, $P$ (Eq.~\ref{eq:IPR}) as a function of the disorder strength $h$. Increasing $h$ leads to an increase (decrease) of $P \, (-\ln(P))$  indicating greater localization. However, for a fixed disorder strength, $h$, $P\,(-\ln(P))$ decreases (increases) with $L$, thereby indicating that the critical disorder strength for the MBL transition diverges in the thermodynamic limit. These results have been obtained from the middle 1/3 eigenstates, and the disorder averaging has been pefromed over 5000 realizations for $N=8$, 10, and 12, and $200$ realizations for $N=14$.} 
\label{fig:Figure6}
\end{figure}

The analysis of the quench dynamics presented in the previous section leads to a natural question: is there a critical value of the disorder strong, $h_c$ beyond which the PWR2 model exhibits many-body localization in thermodynamic limit? We address this question by investigating the spectral statistics and the eigenstate properties of this system.\\
We begin our analysis by determining the spectral statistics. To this end, we compute the disorder-averaged ratio of successive energy gaps~\cite{oganesyan2007localization,atas2013distribution,giraud2022probing}:
\begin{equation}
    r_n=\frac{{\rm min}\{\delta_n,\delta_{n+1}\}}{{\rm max}\{\delta_n,\delta_{n+1}\}},
\end{equation}
where $\delta_n = E_{n+1} - E_{n}$. The chaotic phase is characterized by the Wigner-Dyson statistics ($\langle r \rangle_{\rm WD}\approx 0.5307$), and the localized phase is characterized by Poisson statistics $\langle r \rangle_{\rm P}\approx 0.3863$. Figure~\ref{fig:Figure5}(a) demonstrates that $\langle r \rangle \approx \langle r \rangle_{\rm WD}$ in the weak-disorder regime and it approaches $\langle r \rangle_{\rm P}$ with increasing disorder. However, we note that the $\langle r \rangle$ curves shift to the right with increasing $L$, indicating that the critical disorder strength for many-body localization, $h_c$ increases with system size. We further note that the intersection point, $h_i$ for consecutive $\langle r \rangle$ keeps shifting to larger values with increasing system size. These results indicate that the MBL phase may not exist in the thermodynamic limit.\\
We explore this issue further by systematically investigating the eigenstate entanglement for different values of $h$ and $N$. Our results are reported in Fig.~\ref{fig:Figure6}. We find that in the weak disorder regime, the average value of the half-chain entanglement, $\langle S\rangle$ for mid-spectrum eigenstates is close to the Page value~\cite{page1993average}. Increasing disorder leads to a lowering of the eigenstate entanglement. However, for a fixed disorder strength, $h$, $\langle S\rangle$ grows with the system size indicating the absence of a crtical $h_c$ for the the thermalization-localization transition. We supplement this analysis by computing the variance of $S$, $\delta S=\sqrt{\langle S^2 \rangle- \langle S \rangle^2}$. It is known that in the thermodynamic limit $\delta S = 0$, both in the thermal and many-body localized phase, and it diverges at the transition between the two phases~\cite{kjall2014many}. Thus, for finite-sized systems, the MBL phase transition point corresponds to the peak of $\delta S$. Our results shown in Fig.~\ref{fig:Figure6}(a) demonstrates that this peak shifts with increasing $L$, thereby indicating the absence of the MBL phase in the thermodynamic limit.\\
The results obtained so far strongly indicate the absence of MBL in this model. We provide further evidence for this inference by computing the inverse participation ratio of the eigenstates~\cite{tikhonov2018many}:
\begin{equation}
    P = \sum_{j=1}^{\mathcal{N}} \vert \langle \mu_j \vert \psi_n \rangle\vert^4,
    \label{eq:IPR}
\end{equation}
where $\vert \mu_j \rangle$ denote the computational basis states. The MBL phase is characterized by localized states by $P \sim 1$ and consequently $\ln(P) \sim 0$. On the other hand, the mid-spectrum eigenstates of chaotic many body-systems usually exhibit homogeneous spreading over the entire Hilbert space, leading to $P \sim \mathcal{N}^{-1}$, leading to $-\ln(P) \propto L$. Our results are shown in Fig.~\ref{fig:Figure6}(b). We find that increasing disorder leads to a decrease of $P$; however, $-\ln(P) \propto L$ even in the strong disorder regime. Furthermore for a fixed $h$, $-\ln(P)$ increases with $L$, thereby indicating that $h_c \rightarrow \infty$ in the thermodynamic limit. Our results clearly demonstrate that although strong disorder can induce localization for finite-sized PWR2 spin chain, the critical value for the MBL transition, $h_c$ diverges in the thermodynamic limit.

\section{Conclusion and Outlook}
\label{sec:Conclusion}
In this work, we have analyzed the quench dynamics and spectral properties of the \emph{Power of Two} model in the presence of disorder. In the absence of disorder, this system serves as a powerful platform for realizing fast scrambling and the consequent generation of metrologically useful entangled states. We demonstrate that disorder leads to localization in the system, leading to slow scrambling, initial state memory retention, and low entanglement entropy. Notably, the OTOC, $C_{ij} (t)$ exhibits a non-monotonic dependence on $\vert i - j \vert$, unlike other long-range interacting models; this non-monotonicity originates from the non-local nature of interactions in this system.\\
Furthermore, we have carefully examined the spectral properties of the system and found that strong disorder induces Poissonian statistics, and low values of eigenstate entanglement and Inverse participation ratio. These results establish that for a fixed system size, $L$, there is a critical disorder strength, $h_c$ at which the system undergoes a chaos-MBL transition. Intriguingly, however, $h_c$ increases monotonically with $L$, thereby establishing that this system does not exhibit MBL at finite disorder strength in the thermodynamic limit. Our results establish the robustness of chaos in the PWR2 model. \\
There are several interesting avenues for future work. Firstly, it would be interesting to examine the growth of spin squeezing~\cite{ma2011quantum} and quantum Fisher information~\cite{liu2020quantum} in the disordered PWR2 model. Moreover, it would be intriguing to examine the dynamics of this system under periodic driving, and devise protocols to realize non-ergodic phenomena such as Hilbert space fragmentation~\cite{moudgalya2022hilbert,wampler2023arrested} and dynamical freezing~\cite{bhattacharyya2012transverse,haldar2021dynamical,gangopadhay2025counterdiabatic} in sparse-coupled long-range models. Finally, the dynamics of these systems in the presence of dissipation could be an interesting avenue for future research~\cite{fazio2025many}.\\

\section*{Acknowledgements}
SC acknowledges funding under the Government of India’s National Quantum Mission grant numbered DST/FFT/NQM/QSM/2024/3. We acknowledge the use of the HPC clusters at HRI.

\appendix
\bibliography{refs}
\end{document}